# Towards a Procedure Optimised Steerable Microcatheter for Deep Seated Neurosurgery

Ayhan Aktas *Student, IEEE,*, A. Anil Demircali *Member, IEEE,*, Riccardo Secoli, Burak Temelkuran *Member, IEEE,* and F. Rodriguez y Baena *Member, IEEE,*

*Abstract*—In recent years, the steerable needles have at- tracted significant interest in Minimally Invasive Surgery (MIS). Amongst these, the flexible Programmable-bevel tip needle (PBN) concept has successfully achieved an in-vivo demonstration to evaluate the feasibility of Convection Enhanced Delivery (CED) of chemotherapeutics within the ovine model, with a 2.5 mm PBN prototype. However, further size reduction is necessary for other diagnostic and therapeutic procedures involving deep- seated tissue structures. Since PBNs have a complex cross-section geometry, standard production methods, such as extrusion, fails as the outer diameter is reduced further. This paper presents our first attempt to demonstrate a new manufacturing method for the PBN that employs thermal drawing technology. Experimental characterisation tests were performed for the 2.5 mm PBN and a new 1.3 mm Thermally Drawn (TD) PBN prototype described here. The results show that thermal drawing presents a significant advantage in miniaturising complex needle structures. However, the steering behaviour is affected due to the choice of material in this first attempt, a limitation which will be addressed in future work.

## I. Introduction

Minimally Invasive Surgery (MIS) has undergone significant growth in the last few decades due to its potential advantages for surgical outcomes, such as shorter re- covery rates, reduced hospitalization, and less tissue disruption [1]. Percutaneous interventions with needles are among the most common application areas of MIS [2]. Straight needles are frequently used in MIS neurosurgery procedures such as biopsy, blood sampling fluid delivery/extraction, and tumour ablation [3]. The effectiveness of these interventions depends on the surgical target being reached precisely and accurately without harming healthy tissues (i.e., with minimal damage to the surrounding tissue), which is frequently impaired by instrument design constraints. Since a straight needle may not be able to access a lesion on the first try, dynamic compen- sation for tip misplacement is impossible without retraction or reinsertion [4]; researchers have been developing robotic steerable needles through a range of designs [5]. One particular design, the programmable-bevel tip needle (PBN), has recently been deployed *in vivo* for the first time in order to evaluate the feasibility of Convection Enhanced Delivery (CED) of chemotherapeutics through preferred curvilinear paths that align to specific anisotropic brain structures [2].

The bio-inspired PBN design features multiple bevel-tip segments (generally four segments with multiple lumens per segment) held together through an interlocking mechanism. It uses the relative motion of the segments to generate a dynamic offset at the tip that enables steering during the insertion process, as illustrated in Fig. 1.

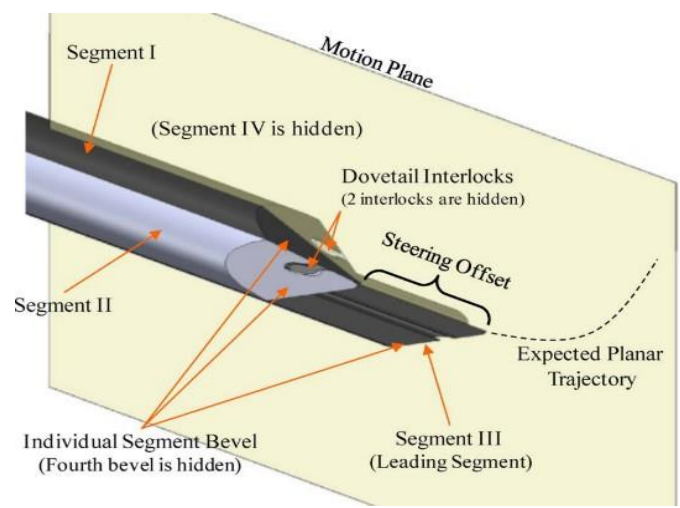

Fig. 1. The Programmable Bevel-tip Needle (PBN) Steering Concept [6]

Starting a decade ago, the first generation of PBN prototypes had a 12 mm diameter, manufactured by rapid prototyping technology, using rubber-like materials such as TangoBlack and VeroWhite [6]. The manufacturing of such complex de- signs has finally reached a clinically viable size of 2.5 mm diameter through a conventional extrusion manufacturing pro- cess [7], the gold standard for medical catheter production. The PBN segments were produced in a coloured bio-compatible polymer, each with two working channels. This version was used as an implantable device during an *in vivo* study on the ovine model, demonstrating the clinical viability of the design for the first time [2]. Though clinically applicable, a further

A. Aktas and F. Rodriguez y Baena are affiliated with Mechatronics in Medicine Laboratory, Department of Mechanical Engineering, Imperial College London, A.A. Demircali and B. Temelkuran are affiliated with the Department of Metabolism, Digestion and Reproduction, Imperial College London, R. Secoli is affiliated with The Hamlyn Centre for Robotic Surgery, Imperial College London, Exhibition Rd, South Kensington, London SW7 2BX.
e-mail: f.rodriguez@imperial.ac.uk
Corresponding author: F. Rodriguez y Baena
The work of Ayhan Aktas was supported by the Republic of Turkiye.

size reduction would be necessary for applications involving deep-seated tissue structures, such as Laser Interstitial Thermal Therapy (LiTT) [8]. Due to the complex design of the PBN, standard manufac- turing methods have limited capacity to achieve further size reduction. Therefore, alternative solutions to decrease the over- all size need to be explored. In this study, we introduce a new manufacturing method for PBNs that employs thermal drawing technology [9], demonstrating the potential to manufacture small-sized complex catheter cross-sectional geometries with unprecedented detail.

Thermal drawing is a manufacturing process that produces longitudinally homogenous fibres while keeping the cross-sectional integrity of the preform. It achieves this by heating a material to a glassy state and drawing it from metres to kilometres-long fibres [10]. Heat and draw tension are applied to elongate the glassy preform section into a fibre. 3D printing technology allowed the creation of the preform with a complex cross-section that could be more expansive and challenging using other approaches, such as molding.

Thermally drawn multi-material fibres with varying geometries, material compositions, and functionalities have been achieved in the past 20 years. This method was previously employed at the micro-nano scale to produce complicated and asymmetrical structures at large scales. The first type of multi-material fibres demonstrating large photonic band gaps dates back to 2002 [11], [12]. Fibre technology involves many different technologies, such as photodetection, thermal sensing, chemical sensing, optical communication, and mi- crofluidics [13]–[16]. In addition, thermal drawing can be used for electronics (e.g., fabric-based and wearable energy-storing systems) and functional textiles [17]. With the advent of additive manufacturing techniques, it became possible to fabricate complex and arbitrary structures at the micrometre scale. Due to these advantages, the thermal drawing technique was employed here to produce each PBN segment, as it allowed miniaturization of the complex cross-section beyond what was possible until now with conventional manufacturing processes. We used a 3D printer (i.e., fused filament fabrication) to create complex geometries with cavities and produced increasingly small prototypes through an iterative refinement process.

This paper describes the fabrication of a sub-millimetre size needle segment by thermally drawing a 3D printed preform (further reducing the needle size by approximately 50%) and comparative characterisation tests against the latest PBN described in [7]. These characterisation tests were performed as described in [18], [19], with refinements to the process that include flexural stiffness and tensile strength assessment. A stereo camera pair was used for curvature estimation of the needle to investigate the relationship between segments offset and curvature.

This paper is organised as follows. Section II provides detailed information about the current needles and the characterisation methods used for mechanical testing and bending performance assessment. The experiments conducted are explained in Section III. In Section IV, experimental results are summarised. Finally, this paper concludes with a discussion of the advantages and disadvantages of each prototype and manufacturing method, with clear implications for the future of these technologies.

## II. Materials and Methods

### A. Catheter Design and Manufacture

*1) Extrusion Manufactured (EM) PBN Catheter:* The EDEN2020 modular robotic environment for precision neurosurgery for the PBN prototype used here is comprehensively described in [2]. The clinical experiments used a 2.5 mm outer diameter 4-segment PBN catheter, as shown in Fig. 2. The needle segments are fabricated by Xograph Healthcare Ltd. via extrusion in medical-grade Poly(vinyl chloride), which has 89 Shore "A" hardness, and colour particles were added to the mix in order to achieve a colour-coded design, where each segment is easily identifiable in the operating theatre (OR). The manufactured PBN segments are nano-coated with Poly(para-xylylene) to ensure biocompatibility and minimal friction between the segments while the segments are sliding with respect to one another. Figure 2 illustrates the bespoke catheter segments and their cross-sectional view. In order to minimise the chance of buckling outside of the gelatin, a trocar with a 2.7 mm inner diameter was used. Each catheter segment contains a male and a female part that interlocks with adjacent segments. Each segment has two 0.3 mm diameter working channels for sensorisation and instrument delivery, and the segment's tips are bevelled at a 45° angle from the neutral axis to aid the insertion process, as described in other studies (e.g. [7], [20]).

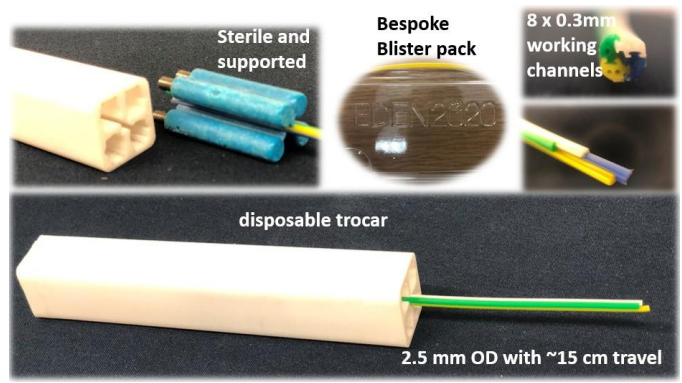

Fig. 2. Extrusion Manufactured (EM) Biocompatible PBN

*2) Thermally Drawn Catheter (TM):* Due to its unequivocal advantages when working at tiny scales, we used Thermal drawing of 3D printed preforms to manufacture each PBN section of a miniaturised PBN design. Standard Computer-Aided Design (CAD) software (SolidWorks, Dassault Systemes, France) was used to construct the 3D printed preform design. Several design strategies were considered during preform creation and iteratively optimised to achieve the intended segment dimensions and sliding performance of the interlocking mechanism by considering the influence of thermal expansion.

A commercially available Ultimaker 3 Extended printer (UltiMaker BV, Netherlands) was used with Poly Carbonate (PC) material (Ultimaker PC Transparent, 2.85 mm filament), which has 82 Shore "D" hardness and a 0.4AA print core

(non-abrasive plastics). The layer thickness was chosen as 0.1 mm with a %100 infill density.

Compared to the default speed of 250 $mm/s$, the print core travel speed was adjusted to 40 $mm/s$ to create smoother surfaces. For transparent PC printing, the following temperatures were taken into account: nozzle temperature $T_n = 270$ $°C$ and bed temperature $T_b = 107 °C$. The dimensions of the 3D printed preform were 40 mm in diameter and 100 mm in length. The central lumen of the preform was 8.6 mm in diameter, and the targeted draw ratio was 40 to reach 0.21 mm central lumen. Feeding the preform into a 3-zone furnace with a speed of $v_f = 2.5\ m/min$ and pulling the fibres with $v_d = 1.4\ m/min$, the 4 cm diameter preform with a 10 cm drawable preform length yields 160 metres of PBN sections. The temperature of the 3-zone furnace used for thermally drawing the fibre was as follows: top zone: 140 $°C$, middle zone: 200 $°C$, and bottom zone: 85 $°C$. The schematic representation of the thermal drawing process can be seen in Fig. 3.

Figure 4 shows the Thermally Drawn (TD) segments and their respective cross-sectional view. In order to minimise the chance of buckling outside of the gelatin phantom, a specially designed trocar with a 1.3mm inner diameter was constructed and employed in the characterisation experiments. Each catheter segment has a 1x0.21 mm diameter working channel for sensorisation and clinical applications, and the segment's tips are bevelled at an angle of 75° from the neutral axis.

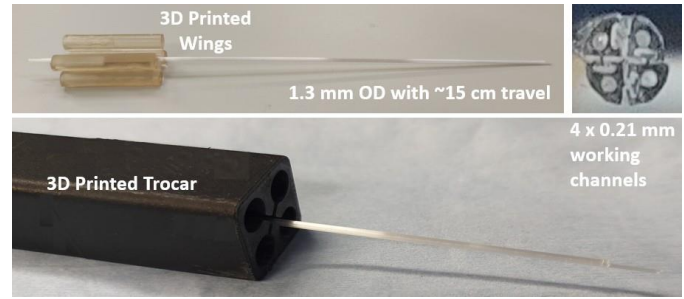

Fig. 4. Thermally Drawn (TD) Programmable Bevel Tip Needle

### B. Characterization Methods

*1) Mechanical Feature Testings:* Structural tests were performed to measure the material characteristics of the two PBN designs used in our *in vivo* work and the thermally drawn prototype described here. A flexural rigidity test measures the needle sensitivity to medium changes. This sensitivity can either complement the needle steering or act as a competing mechanism that works against it, depending on the steering strategy of the needle [21]. Consequently, three-point bending tests were conducted to measure the flexural stiffness of the EM PBN and thermally drawn segments with a 30 mm specimen length. The test was performed with a 10N Instron Load cell (INSTRON, US) with a cross-head velocity of 1 mm/min.

In addition, tensile tests were conducted to identify the tensile stress-strain behaviour of each PBN. The tests were performed using the same samples with a 1 kN Instron load cell with a 50mm/min pulling velocity. Lastly, the holding capacity of the interlocking mechanism was investigated, as unwanted separation of the segments would lead to failure of the PBN. A tensile test was employed to measure each prototype's strength of the interlocking mechanism. The test is performed on two interlocked segments fixed together through a holder attached to the wings. The two segments are subsequently pulled out from one another with a 1mm/min velocity. The force at the disconnection was identified by a sudden decreasing point in tensile force and was taken as the breakout force. Each experiment is repeated three times. The test setups are shown in Fig. 5.

*2) Curvature Estimation Method:* The 3D curvatures of each needle were collected using a stereo camera setup, following camera calibration, shape digitization and shape reconstruction implemented using the R package StereoMorph [22]. The stereo camera setup consists of two fixed C920 HD Pro cameras (Logitech inc.) with fixed focal lengths

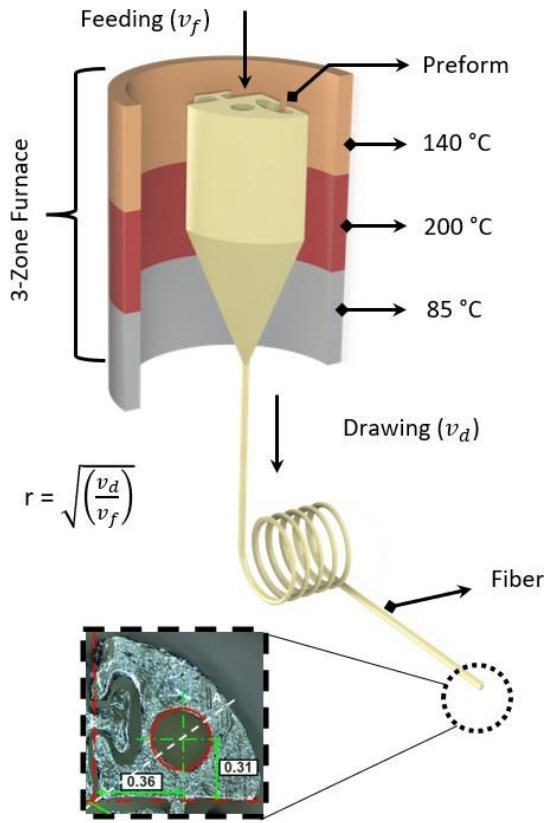

Fig. 3. Thermal Drawing Process with 3D Printed Preform

The thermal drawing process produced a long PBN-shaped fibre segment with a radius of 0.65 mm, and afterwards, the long fibre segment was trimmed to the desired length of 200 mm, which was chosen to match the existing prototype described in [20]. As in many previous works, each short segment was fixed to 3D-printed "wings" used to assemble each to an actuation mechanism, and needle assembly was completed manually by interlocking four individual segments.

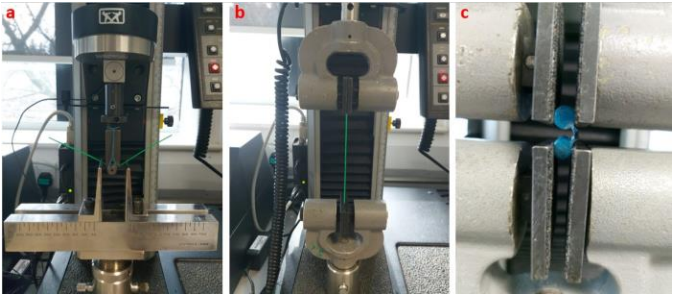

Fig. 5. a)Flextural Stiffness Test, b) Tensile Strength Test, c) Interlocking Breakout Force Test

and overlapping fields of view, calibrated using a classic checkerboard pattern. The calibration of two stereo cameras has been done using the StereoMorph's calibration process as described in [22]. Following each insertion, the needle is photographed using the stereo cameras, and the needle shape is manually digitized in each camera view using the StereoMorph digitizing application. We used 32 evenly spaced points to capture the curvature of each shape. Fig 6 shows an example of the digitizing tool used for one of the experiments. The purple dots show the selected digitization points from the right side camera, and selected points appear on the right-hand side of the program. In addition, the application shows the epipolar lines from the base and tip of the needle to match the left and right camera images. Following point selection, the digitization points are interpreted using the Bezier Curve approach, which is embedded in the application. The digitized curvatures were then reconstructed from base to tip into 3D, according to the calibration coefficients identified offline. Finally, refractive index correction is taken into account to compensate for the measurements obtained from the gelatin medium.

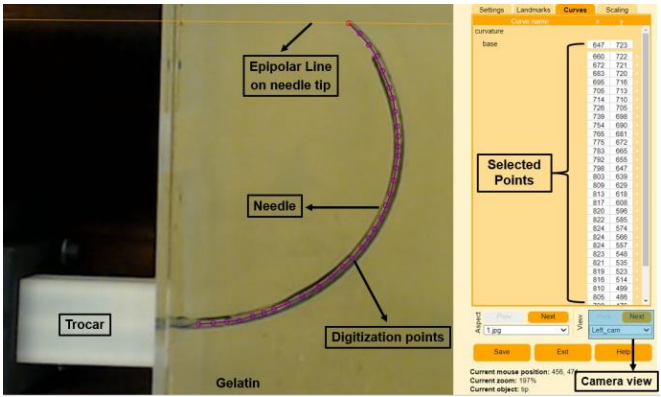

Fig. 6. StereoMorph Digitizing Application from the Right Camera View

## III. EXPERIMENTAL VALIDATION

The steering characteristics of both PBNs were assessed with a series of experimental needle insertions. These were performed using the same technique as in earlier investigations [6], [19], [23]. Fig. 7 shows a diagram of the needle steer- ing setup for the characterisation tests. The insertions were performed using four linear actuators, controlled via software developed in-house [24]. Each linear actuator is connected to a needle segment's wing via a nitinol rod (which allows sliding between the segments), travelling at a fixed speed of 1 mm/s, as in previous studies [19], [6], [18]. Each segment is enclosed within a 3D-printed trocar core to preserve the needle segments' alignment before insertion. The experimental setup is shown in Fig 8. The maximum achievable curvature for a range of offsets was used to compare the needles' steering capabilities, including the minimum achievable radius of curvature [19]. The characterisation tests were conducted in a temperature-controlled environment at (20-21) °C using a 6% by wight bovine gelatin phantom (Chef William Powdered Gelatin) [18], which is an acceptable first-order approximation to human brain white matter [25].

All needle segments were initially aligned and inserted 20 mm into the phantom. The furthest extended segment or segments were kept fixed, and the remaining segments were driven back to achieve specified offset configurations. Then, all four linear actuators were synchronously driven to achieve insertion of 110 mm as the in previous characterisation works [19].

Insertions were performed with the configurations of a single segment forward and two segments forward to evalu- ate the achievable curvature performance in all four planes, with increasing insertion offsets of 5, 10, 15 and 20 mm. A minimum of 10 insertions were carried out for each tip configuration. The achieved needle trajectories were measured using the calibrated stereo camera pair after each insertion.

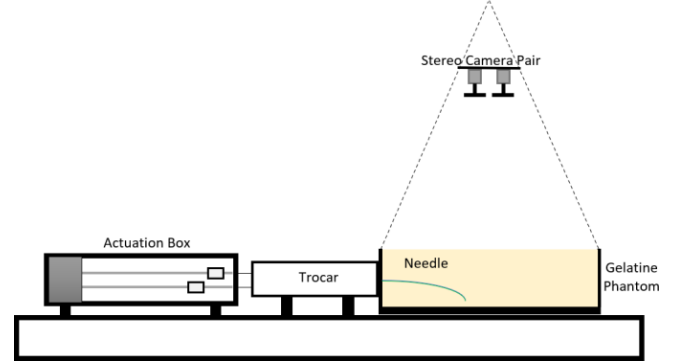

Fig. 7. Diagram of the Experimental Setup

Calibration of the stereo cameras was done with an 8x11 checkerboard and StereoMorph's digitising app resulting in a 0.423 pixel (px) epipolar mean error. At the end of each insertion, we manually identified the curvature projections for the stereo camera images, to be reconstructed in 3D using the StereoMorph digitising tool, as in Fig 6. StereoMorph reconstruction mode is used after point selection to obtain needle curvatures in 3D. In order to account for the refraction at the air-gelatin interface, a universal algorithm [26] was used because the stereo camera calibration was performed in the free air medium, whereas the experimental trials were conducted in a gelatin medium.

Following refraction correction, the curvature vector $\kappa_{exp}$ was estimated from the resulting trajectories. Murthy's 3D circle fit function [27] was used to estimate the best-fit radius (R) of each curvature from which the magnitude of the curvature could be computed as $\kappa = 1/R$. The steering and horizontal planes (where the needle is originally placed) were used to calculate the insertion angle ($\beta$). The resulting curvature vector for each insertion was calculated as in [19].

$$\mathcal{K}_{exp} = \mathcal{K} \begin{bmatrix} \cos(\beta) \\ \sin(\beta) \end{bmatrix}$$

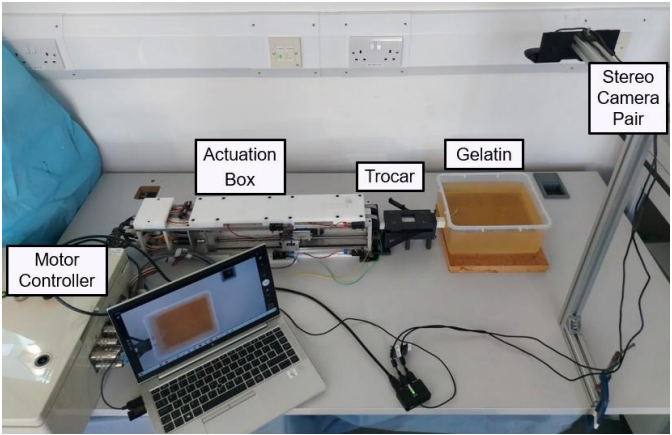

Fig. 8. Experimental Setup

## IV. RESULTS

### A. Mechanical Features

Fig. 9 shows the cross-section comparison of the Extrusion Manufactured (EM) PBN and thermally drawn (TD) PBN segments, viewed under a microscope. The findings indicate that this method reduced the diameter to about half of the state-of-the-art size while retaining full functionality and four working channels. We opted to half the number of lumens per segment in order to maintain a 200-micron working channel in the smaller PBN for eventual integration with our fibre brag grating-based experimental setup [24].

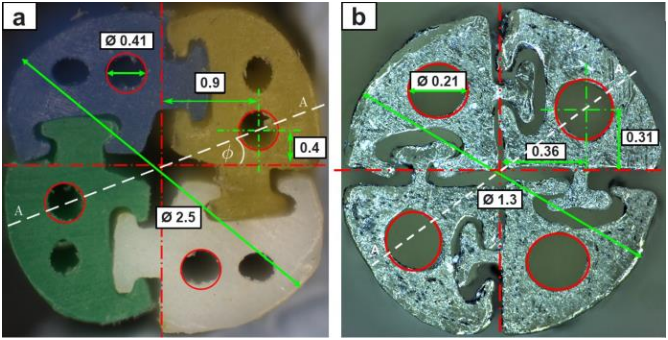

Fig. 9. a) EM Needle, b) TD Needle

The mean flexural stiffness, tensile stress, and interlocking mechanism breakout force values for the EM and TD PBN segments (2.5 mm and 1.3mm diameter) are given in Table 1. Segment rigidity strongly influences the curvature performance during the insertion stages of the needle. The thermally drawn material has higher stiffness than the plastic employed with the extrusion method. As a result, the catheter curvature performance worsened despite the smaller diameter, while the interlocking strength was improved. Specifically, with the 1.3 mm prototype, the effect on steering performance was 30% worse, while the interlocking strength was 50% better. Qualitatively, the thermally drawn segments work more efficiently, moving with respect to one another more fluidly and predictably.

TABLE I
MECHANICAL FEATURES

|  | *Mean Flexural Stiffness (N/mm)* | *Mean Tensile Stress (MPa)* | *Mean Interlocking Breakout Force (N)* |
|---|---|---|---|
| *2.5mm EM* | 0.023 | 13.10 | 5.47 |
| *2.5mm TD* | 0.38 | 53.66 | 18.52 |
| *1.3mm TD* | 0.031 | 51.45 | 10.94 |

### B. Offset vs Curvature

Two assess the steering behaviour of the new prototype against our state-of-the-art preclinical PBN; two experiments were performed: a "single leading segment" and "two leading segments" insertions, using the EM and TD PBNs. Fig. 10 and Fig. 11 display the mean value and standard deviation for ten experiments in each test. For each, a linear fit of the curvature data in relation to steering offset is also depicted.

Fig.10 displays the offset-curvature characteristics for a single-segment insertion of both needles. The minor disparity between the curvature values for positive and negative offsets may be caused by needle torsion, positioning uncertainty during the insertion process, and deformation. The maximum curvature achieved for the EM PBN is $0.0242 \pm 0.003 mm^{-1}$, which corresponds to a radius of curvature of $41.307 mm$. The resulting curvature of this needle is substantially higher than the $0.0192 mm^{-1}$ previously reported in [19], under similar settings. This result is expected due to the use of a less stiff material, resulting from the addition of colour pigments and the addition of a working channel per segment (the PBN in [19] only possessed one working channel per segment). The maximum curvature achieved for the TD PBN is $0.0092 \pm 0.001 mm^{-1}$, which corresponds to a radius of curvature of $109.113 mm$. Table II shows the experimental results for single-segment insertion. The first column shows the offsets for each needle, and the last three columns of Table II report on the mean curvature ($\kappa$ [mm]), the radius of curvature (R [mm]) and angle of insertion ($\beta$).

The same experiments were conducted with two forward segments in Fig. 11. The maximum curvature achieved for the EM PBN with two forward segments is $0.0124 \pm 0.001 mm^{-1}$, which corresponds to a radius of curvature of $82.644 mm$. The maximum curvature achieved for the TD PBN with two forward segments is $0.0069 \pm 0.001 mm^{-1}$, which corresponds to a radius of curvature of $144.921 mm$. Table III shows the experimental results for two-segment insertions. The first column shows the offsets for each catheter, and the last two columns of Table III report on the mean curvature ($\kappa$ [mm]) and radius of curvature (R [mm]). The insertion angle is not reported because the curvatures here are measured in 2D, as they were performed in the plane normal to the camera line of sight. The reason for the difference between the single segment and two segments is discussed in our previous study [18]. It relates to the effective stiffness of the programmable bevel achieved with two segments compared to just one. In Fig. 10 and Fig. 11, there is a linear relationship between curvature and offset, and we can see this in both needle types.

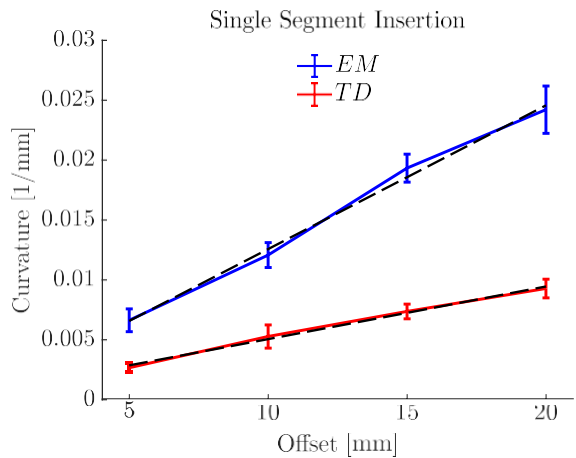

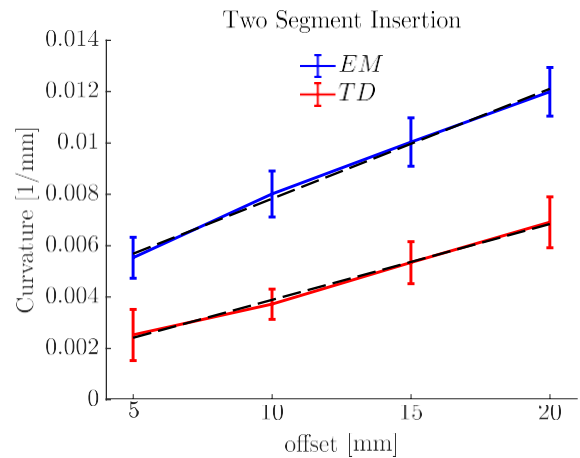

Fig. 10. Single Forward Segment Insertion Results (Offset-Curvature)

Fig. 11. Two Forward Segments Insertion Results (Offset-Curvature)

TABLE II
SINGLE SEGMENT INSERTION OFFSET-CURVATURE RELATIONSHIP

|  | Offsets (mm) | Mean $\kappa$ (1/mm) | Mean R (mm) | Mean $\beta$ (degree) |
|---|---|---|---|---|
| EM PBN | 5 | 0.0066 | 151.488 | 28.89 |
|  | 10 | 0.0120 | 83.306 | 42.68 |
|  | 15 | 0.0193 | 52.454 | 54.58 |
|  | 20 | 0.0242 | 41.307 | 69.20 |
| TD PBN | 5 | 0.0026 | 385.516 | 8.93 |
|  | 10 | 0.0052 | 192.432 | 19.24 |
|  | 15 | 0.0073 | 136.410 | 28.12 |
|  | 20 | 0.0092 | 109.113 | 34.30 |

TABLE III
TWO FORWARD SEGMENTS INSERTION OFFSET-CURVATURE RELATIONSHIP

|  | Offsets (mm) | Mean $\kappa$ (1/mm) | Mean R (mm) |
|---|---|---|---|
| EM PBN | 5 | 0.0055 | 181.488 |
|  | 10 | 0.0080 | 125.036 |
|  | 15 | 0.0102 | 98.034 |
|  | 20 | 0.0121 | 82.644 |
| TD PBN | 5 | 0.0025 | 400.056 |
|  | 10 | 0.0037 | 270.270 |
|  | 15 | 0.0053 | 188.679 |
|  | 20 | 0.0069 | 144.921 |

## V. Discussion

This study presents an alternative technique to manufacture complex-shaped needles and catheters, the size of which exceeds conventional extrusion limits. With thermal drawing, it has been shown that it is possible to produce sub-millimetre segments with high tolerances and excellent shape control. Considering the length of time to miniaturise each successive PBN prototype (Fig.12) using conventional manufacturing techniques, it took over eight years to reduce the size of our very first 12 mm prototype down to 2.5 mm. However, the thermal drawing method allowed us to achieve a 50 % reduction over three months, an achievement that will now enable the application of the steerable system described in [2] to a broader range of diagnostic and therapeutic interventions.

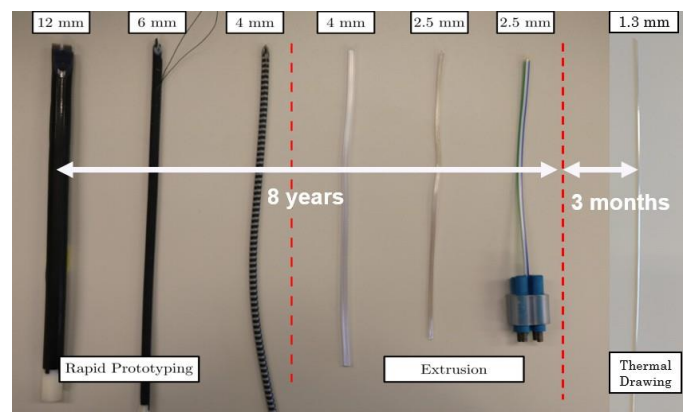

Fig. 12. Size Change of PBN Over Time [20]

Inaccuracies in the interlocking mechanism exist, but they do not affect the sliding behaviour or breakout strength be- tween the segments, confirming the use of thermal drawing as a substitute for extrusion. Additionally, when comparing the thermal drawing method to extrusion, the former does not need additional post-production processing (e.g., nano-coating to improve the sliding behaviour), representing a marked advantage. Besides, the interlocking strength of the TD catheter is 50 % stronger, meaning segment separation will not be a problem as we reduce the outer diameter further, and the sliding behaviour of the TD segments is markedly better.

On the other hand, curvature characterization results show that the new TD PBN performance steers less than the EM PBN despite almost halving the needle diameter. This demonstrates the importance of material stiffness on curvature performance. TD PBN steering performance is low because the material (Poly Carbonate) used for thermal drawing has higher stiffness compared to EM PBN material Poly(vinyl chloride). PC material was chosen for this study due to its availability and local experience with the manufacturing technique. Refining the stiffness profile with softer materials and tuning the thermal drawing process will be a focus of future work, enabling us to match and exceed steering performance. The TD needle should also be less disruptive during the insertion process, as the smaller outer diameter will impact the size of the track left behind. Visual inspection showed that the TD needle significantly reduced damage at the entrance and along the needle track. However, in future work, quantitative analysis of the same will be carried out, as in Leibenger's study [28], to confirm this conclusively.

## VI. Conclusion and Future Work

This paper describes ongoing research into using the thermal drawing method to produce complex-shaped catheters at the sub-millimetre size. Specifically, it demonstrates its application to needle steering, with the design, manufacture and characterisation of a 1.3 mm Programmable Bevel-tip Needle able to meet the size requirements for deep-seated neurosurgical interventions. Mechanical tests were performed to compare the new prototype against our state-of-the-art pre-clinical system, manufactured commercially via conventional extrusion. Additionally, fixed-offset experimental insertions on a phantom were performed to compare and contrast the steering capabilities of the new PBN. The main results show that thermal drawing resulted in a 50 per cent reduction in outer diameter (from 2.5 mm to 1.3 mm) but shallower steering behaviour due to the choice of material employed during the thermal drawing process.

Even though these first experimental findings are encouraging, a thorough examination of the effect of this change on the curvature performance of the needle will be part of future work. We are currently exploring different materials and preform molding methods that will enable the needle to achieve tighter bends. We will also integrate the newly developed catheter within EDEN2020's ecosystem for precision neurosurgery, with the aim to eventually apply our needle steering technology to LiTT. Finally, we will continue to research new potential control schemes, haptic teleoperation, and surgical training techniques.

## Acknowledgments

The authors would like to thank Stephen Laws for his helpful advice on image processing and constructive criticism of the manuscript.

## References

[1] E. P. Westebring-van der Putten, R. H. Goossens, J. J. Jakimowicz, and J. Dankelman, "Haptics in minimally invasive surgery–a review," *Minimally Invasive Therapy & Allied Technologies*, vol. 17, no. 1, pp. 3–16, 2008.

[2] R. Secoli, E. Matheson, M. Pinzi, S. Galvan, A. Donder, T. Watts, M. Riva, D. D. Zani, L. Bello, and F. Rodriguez y Baena, "Modular robotic platform for precision neurosurgery with a bio-inspired needle: System overview and first in-vivo deployment," *PLOS ONE*, vol. 17, no. 10, pp. 1–29, 10 2022. [Online]. Available: https://doi.org/10.1371/journal.pone.0275686

[3] P. Kulkarni, S. Sikander, P. Biswas, S. Frawley, and S.-E. Song, "Review of robotic needle guide systems for percutaneous intervention," *Annals of biomedical engineering*, vol. 47, no. 12, pp. 2489–2513, 2019.

[4] N. K. Patel, P. Plaha, and S. S. Gill, "Magnetic resonance imaging-directed method for functional neurosurgery using implantable guide tubes," *Operative Neurosurgery*, vol. 61, no. suppl_5, pp. ONS358–ONS366, 2007.

[5] N. J. van de Berg, D. J. van Gerwen, J. Dankelman, and J. J. van den Dobbelsteen, "Design choices in needle steering; a review," *IEEE/ASME Transactions on Mechatronics*, vol. 20, no. 5, pp. 2172–2183, Oct 2015.

[6] S. Y. Ko, L. Frasson, and F. R. y Baena, "Closed-loop planar motion control of a steerable probe with a "programmable bevel" inspired by nature," *IEEE Transactions on Robotics*, vol. 27, no. 5, pp. 970–983, 2011.

[7] E. Matheson, T. Watts, R. Secoli, and F. R. y Baena, "Cyclic motion control for programmable bevel-tip needle 3d steering: A simulation study," in *2018 IEEE International Conference on Robotics and Biomimetics (ROBIO)*. IEEE, 2018, pp. 444–449.

[8] M. Pinzi, B. Hwang, V. N. Vakharia, J. S. Duncan, F. Rodriguez y Baena, and W. S. Anderson, "Computer assisted planning for curved laser interstitial thermal therapy," *IEEE Transactions on Biomedical Engineering*, 2021.

[9] R. Chen, A. Canales, and P. Anikeeva, "Neural recording and modulation technologies," *Nature Reviews Materials*, vol. 2, no. 2, pp. 1–16, 2017.

[10] L. van der Elst, C. F. de Lima, M. G. Kurtoglu, V. N. Koraganji, M. Zheng, and A. Gumennik, "3d printing in fiber-device technology," *Advanced Fiber Materials*, pp. 1–17, 2021.

[11] S. D. Hart, G. R. Maskaly, B. Temelkuran, P. H. Prideaux, J. D. Joannopoulos, and Y. Fink, "External reflection from omnidirectional dielectric mirror fibers," *Science*, vol. 296, no. 5567, pp. 510–513, 2002.

[12] B. Temelkuran, S. D. Hart, G. Benoit, J. D. Joannopoulos, and Y. Fink, "Wavelength-scalable hollow optical fibres with large photonic bandgaps for co2 laser transmission," *Nature*, vol. 420, no. 6916, pp. 650–653, 2002.

[13] F. Sorin, A. F. Abouraddy, N. Orf, O. Shapira, J. Viens, J. Arnold, J. D. Joannopoulos, and Y. Fink, "Multimaterial photodetecting fibers: a geometric and structural study," *Advanced Materials*, vol. 19, no. 22, pp. 3872–3877, 2007.

[14] M. Bayindir, A. F. Abouraddy, J. Arnold, J. D. Joannopoulos, and Y. Fink, "Thermal-sensing fiber devices by multimaterial codrawing," *Advanced Materials*, vol. 18, no. 7, pp. 845–849, 2006.

[15] A. Gumennik, A. M. Stolyarov, B. R. Schell, C. Hou, G. Lestoquoy, F. Sorin, W. McDaniel, A. Rose, J. D. Joannopoulos, and Y. Fink, "All-in-fiber chemical sensing," *Advanced Materials*, vol. 24, no. 45, pp. 6005–6009, 2012.

[16] W. Yan, A. Page, T. Nguyen-Dang, Y. Qu, F. Sordo, L. Wei, and F. Sorin, "Advanced multimaterial electronic and optoelectronic fibers and textiles," *Advanced materials*, vol. 31, no. 1, p. 1802368, 2019.

[17] T. Khudiyev, J. T. Lee, J. R. Cox, E. Argentieri, G. Loke, R. Yuan, G. H. Noel, R. Tatara, Y. Yu, F. Logan *et al.*, "100 m long thermally drawn supercapacitor fibers with applications to 3d printing and textiles," *Advanced Materials*, vol. 32, no. 49, p. 2004971, 2020.

[18] C. Burrows, R. Secoli, and F. R. y Baena, "Experimental characterisation of a biologically inspired 3d steering needle," in *2013 13th international conference on control, automation and systems (ICCAS 2013)*. IEEE, 2013, pp. 1252–1257.

[19] T. Watts, R. Secoli, and F. R. y Baena, "A mechanics-based model for 3-d steering of programmable bevel-tip needles," *IEEE Transactions on Robotics*, vol. 35, no. 2, pp. 371–386, 2018.

[20] V. Virdyawan, "Sensorisation of a novel biologically inspired flexible needle," Ph.D. dissertation, Imperial College London, 2018.

[21] N. J. van de Berg, D. J. van Gerwen, J. Dankelman, and J. J. van den Dobbelsteen, "Design choices in needle steering—a review," *IEEE/ASME Transactions on Mechatronics*, vol. 20, no. 5, pp. 2172–2183, 2014.

[22] A. M. Olsen and M. W. Westneat, "Stereomorph: An r package for the collection of 3d landmarks and curves using a stereo camera set-up," *Methods in Ecology and Evolution*, vol. 6, no. 3, pp. 351–356, 2015.

[23] S. Y. Ko, B. L. Davies, and F. R. y Baena, "Two-dimensional needle steering with a "programmable bevel" inspired by nature: Modeling preliminaries," in *2010 IEEE/RSJ International Conference on Intelligent Robots and Systems*. IEEE, 2010, pp. 2319–2324.

[24] A. Donder and F. R. y Baena, "Kalman-filter-based, dynamic 3-d shape reconstruction for steerable needles with fiber bragg gratings in multicore fibers," *IEEE Transactions on Robotics*, vol. 38, no. 4, pp. 2262–2275, 2021.

[25] F. Giulia, "The mechanics of human brain tissue, modelling, preservation and control of materials and structures," 2016.

[26] B. Cao, R. Deng, and S. Zhu, "Universal algorithm for water depth refraction correction in through-water stereo remote sensing," *International Journal of Applied Earth Observation and Geoinformation*, vol. 91, p. 102108, 2020.

[27] S. Murthy, "Sam murthy (2022). best fit 3d circle to a set of points (https://www.mathworks.com/matlabcentral/fileexchange/55304- best-fit-3d-circle-to-a-set-of-points), matlab central file exchange. re- trieved," 2022.

[28] A. Leibinger, A. E. Forte, Z. Tan, M. J. Oldfield, F. Beyrau, D. Dini, and F. Rodriguez y Baena, "Soft tissue phantoms for realistic needle insertion: a comparative study," *Annals of biomedical engineering*, vol. 44, no. 8, pp. 2442–2452, 2016.